\documentclass[10pt,letterpaper,twocolumn]{article} 

\usepackage{ol2}
\usepackage[draft]{hyperref}
\usepackage{amsmath}

\begin{document}

\twocolumn[ 

\title{Directionality of THz emission from photoinduced gas plasmas}


\author{C. K\"ohler$^{1}$, E. Cabrera-Granado$^{1}$, I. Babushkin$^{2}$, L. Berg\'e$^3$, J. Herrmann$^{4}$, S. Skupin$^{1,5}$}

\address{
$^{1}$Max Planck Institute for the Physics of Complex Systems, 01187 Dresden,
Germany\\
$^{2}$Weierstra\ss -Institut f\"ur Angewandte Analysis und Stochastik, 10117
Berlin, Germany\\
$^{3}$CEA-DAM, DIF, F-91297 Arpajon, France\\
$^{4}$Max-Born-Institut f\"ur Nichtlineare Optik und Kurzzeitspektroskopie, 12489
Berlin, Germany\\
$^{5}$Friedrich Schiller University, Institute of Condensed Matter Theory and
Optics, 07742 Jena, Germany\\
$^*$Corresponding author: koehler@pks.mpg.de
}

\begin{abstract}Forward and backward THz emission by ionizing two-color laser pulses in gas is investigated by means of a simple semi-analytical model based on Jefimenko's equation and rigorous Maxwell simulations in one and two dimensions. We find the emission in backward direction having a much smaller spectral bandwidth than in forward direction and explain this by interference effects. Forward THz radiation is generated predominantly at the ionization front and thus 
almost not affected by the opacity of the plasma, in excellent agreement with results obtained from a unidirectional pulse propagation model.
\end{abstract}

\ocis{260.5210, 300.6270, 320.7120, 350.5400}

 ] 

Generation of radiation in the THz range and controlling its spectrum and direction of emission is crucial for applications reaching from nonlinear THz spectroscopy to biomedical and security imaging. Several methods for THz generation, such as photoconductive switches or optical rectification in second order nonlinear crystals are limited in achievable THz field amplitudes due to saturation or material damage for high input intensities. An alternative setup~\cite{cook00,kim07, bartel05, dai06, thomson07, babushkin10a}, where an ionizing two-color laser pulse is focused into a gas cell has attracted much interest, since the obtained THz pulses are characterized by comparably high amplitudes and broad spectra at the same time. The THz emission can be explained by a low frequency component in the plasma current caused by the asymmetric two-color laser field~\cite{kim07}, where the stepwise modulation of the electron current by two-color tunneling ionization plays a crucial role~\cite{babushkin10a}. Within this model, good agreement of forward (FW) emitted THz radiation in experiment and simulation was obtained recently~\cite{babushkin10a}. However, to the best of our knowledge, an investigation of possible backward (BW) THz emission due to this mechanism is still missing.

In this Letter we present a combined analytical and
numerical study of THz generation by ionizing two-color femtosecond pulses in a gas.
We reveal the governing effect for differences in FW and BW emitted THz spectra to be interference of signals from different spatial positions in the plasma channel. Furthermore, we find that opacity of the plasma is of minor influence on FW THz emission in focused pump geometries.

\begin{figure}[htb]
\includegraphics[width=\columnwidth]{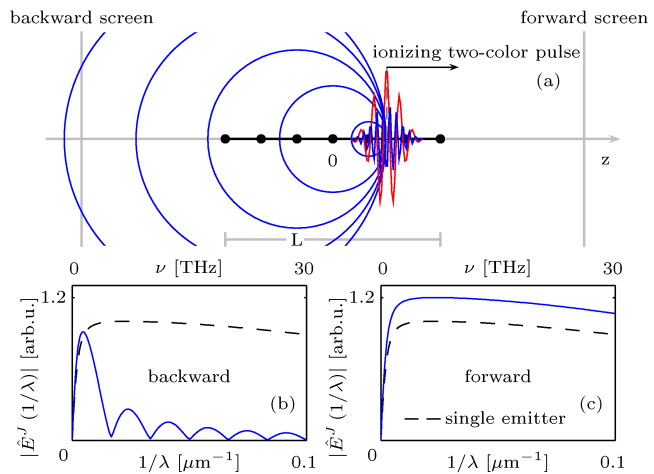}
\caption{\label{fig:huiygensscheme} (a) Schematic illustration of the interference being responsible for shaping FW and BW emission. A plasma line-source of length $L$ created by a propagating two-color pump pulse emits radiation. Blue circles centered around exemplary point-sources represent planes of constant phase of elementary spherical waves. In FW direction (c), spherical waves interfere constructively for all wavelengths and the resulting on-axis spectrum is proportional to the single emitter spectrum. In BW direction (b), the spectral form-factor Eq.~(\ref{eqn:Spectshape}) depletes wavelengths smaller than the source length (here $L=30~\mu$m).}
\end{figure}

Let us first develop a simple picture to explain the main difference between FW and BW emitted fields. A simplified though generic setup is shown in Fig.~\ref{fig:huiygensscheme}. We assume a plasma line-source of length $L$ on the z-axis, which  is meant to be produced by a laser pulse propagating in positive $z$-direction (FW). At each point, the pump pulse ionizes the medium, the generated free electrons are accelerated in the laser field and thus build up a current $J(t)$, which in turn emits electromagnetic radiation.  Due to the propagation of the ionizing pump pulse, points with smaller $z$-coordinates emit radiation earlier than the ones with larger $z$-coordinates. If all propagation effects (diffraction, dispersion, etc.) are neglected, the pulse moves unchanged and all constituent points along the plasma line emit the same field, but shifted in time. This is illustrated in Fig.~\ref{fig:huiygensscheme}(a), where a snapshot of the emitting plasma line is shown. Each blue circle is centered around a different point-source, representing a plane of constant phase of an elementary spherical wave for one wavelength~$\lambda$. To record the resulting spectra on a screen located before (BW) or behind (FW) the line, Jefimenkos's equation~\cite{Jefimenko}
\begin{equation}
 E^J\left(\vec{r},t\right)=-\frac{1}{4\pi\epsilon_0}\int\left(\frac{1}{c^2R}\frac{\partial J\left(\vec{r}',t-R/c\right)}{\partial t}\right)d^3\vec{r}'
\label{eq:jefimenko}
\end{equation}
is used to calculate the electric field emitted from a given current distribution. Here, $R=\lvert \vec{r}-\vec{r}'\rvert$ is the distance between the emitting point-source and the screen.

In this simplified situation, the FW on-axis spectrum is just proportional to the spectrum of a single point emitter $|\hat{E}^J_0(\lambda^{-1})|$, because contributions from emitters along the plasma line add up constructively for all wavelengths~$\lambda$ 
[see Fig.~\ref{fig:huiygensscheme}(c)]. 
In contrast, the BW on-axis spectrum differs significantly from the single emitter spectrum [see Fig.~\ref{fig:huiygensscheme}(b)], since radiation from different emitters is superposed with different temporal delays. The influence of these delays on the BW spectrum can be accounted for by a wavelength dependent form-factor
\begin{equation}
f(\lambda^{-1})=\textrm{Ei}(\frac{i4\pi}{\lambda}(z-L/2))-\textrm{Ei}(\frac{i4\pi}{\lambda}(z+L/2)),
\label{eqn:Spectshape}
\end{equation}
which can be obtained directly from Eq.~(\ref{eq:jefimenko}) by assuming a uniform current density being temporally shifted by $z'/c$ along the line. Here, $\textrm{Ei}(x)=\int_{-\infty}^x \frac{\exp{(x^{\prime})}}{x^{\prime}}dx^{\prime}$ is the exponential integral and $z$ is the position of the screen. The on-axis BW spectrum is then proportional to  $|f(\lambda^{-1})\hat{E}^J_0(\lambda^{-1})|$. Figure~\ref{fig:huiygensscheme}(b)  reveals that the BW spectrum gets depleted for wavelengths $\lambda \ll L$, whereas it coincides with the FW spectrum for $\lambda \gg L$, because for these wavelengths the plasma line appears as a point source. 

\begin{figure}
\includegraphics[width=\columnwidth]{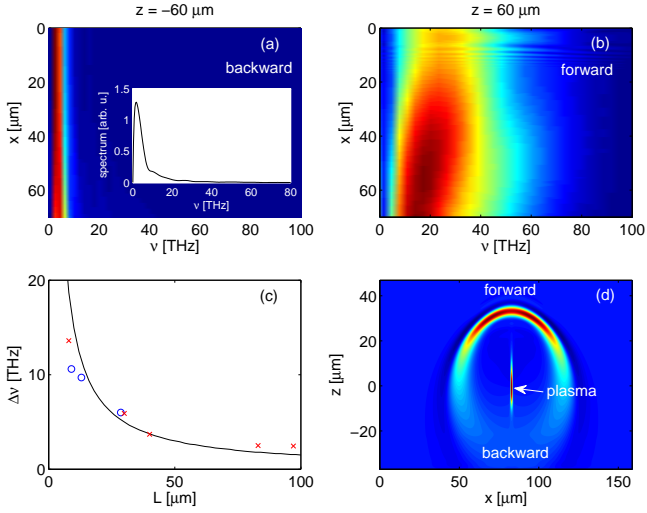}
 \caption{\label{fig:2DSims} 2D FDTD simulations: THz spectra of (a) BW and (b) FW emitted fields. The inset shows the on-axis BW spectrum obtained from a linearly propagated Gaussian pump pulse and Eq.~(\ref{eq:jefimenko}) (3D Jefimenko approach, see text). (c) On-axis BW spectral width vs.\ plasma channel length $L$ from Eq.~(\ref{eqn:Spectshape}) (solid line), 2D FDTD calculations (blue circles), and 3D Jefimenko approach (red crosses). (d) Snapshot of the emitted THz fields ($<100$~THz) and the plasma channel, illustrating the strong (weak) emission in FW (BW) direction.}
\end{figure}

We now confront our predictions with rigorous two-dimensional (2D) Maxwell simulations. We consider two-color input pump pulses 
\begin{align}
 E_{\rm in}(r_{\perp},t)=&\left[\sqrt{1-\xi}\cos\left(\omega_0t\right)+\sqrt{\xi}\cos\left(2\omega_0t+\varphi\right)\right]\nonumber\\
      & \times A\exp\left(-\frac{r_{\perp}^2}{w^2}-\frac{t^2}{\sigma_t^2}\right)
\end{align}
with amplitude $A$, beam width $w$, pulse duration $\sigma_t$, $r_{\perp}=x$ (2D) or $r_{\perp}=\sqrt{x^2+y^2}$ (3D), relative strength of fundamental ($\omega_0=2\pi\nu_0$, $\nu_0=375$~THz) and second harmonic $\xi$, relative phase $\varphi$, being focused into argon gas at atmospheric pressure. The density of generated free electrons obeys
$\partial_t{\rho}_e=W_{\rm ST}(E)\left[\rho_{at}-\rho_e(t)\right]$, where $\rho_{at}$ is the neutral atomic density and 
$W_{\rm ST}(E)$ a field dependent tunneling ionization rate~\cite{thomson07,babushkin10a}. Further on, we assume zero velocity for newly born electrons which are then accelerated in the electric field, leading to the current density \cite{thomson07,kim07, babushkin10}
\begin{equation}
 \frac{\partial}{\partial t}{J_e}+\frac{1}{\tau_c} J_e = \frac{q^2}{m_e}E(t) \rho_e(t).
\end{equation}
We model 2D transversal electric (TE) field evolution by means of the finite difference time domain (FDTD) method~\cite{Joseph_1997}, solving Maxwell's equations
\begin{equation}
\begin{split}
 \mu_0 \frac{\partial H_x}{\partial t} =& \frac{\partial E_y}{\partial z},\qquad -\mu_0 \frac{\partial H_z}{\partial t} = \frac{\partial E_y}{\partial x},\\
\frac{\partial D_y}{\partial t} +J_e =& \frac{\partial H_x}{\partial z}-\frac{\partial H_z}{\partial x}.
\end{split}
\label{eq:fdtd}
\end{equation}
Linear dispersive properties of argon are included via $\hat{D}_y(\omega)=\epsilon_0 n^2(\omega) \hat{E}_y(\omega)$, where we use the refractive index $n(\omega)$ given in \cite{dalgarno60}. In all FDTD calculations we neglect nonlinear contributions to the polarization, because they play a minor role for THz generation in the present configuration.
Figure~\ref{fig:2DSims} shows results for a focused ($f=125~\mu$m) pump pulse with $A=13.6$~GV/m, $w=32~\mu$m, $\sigma_t=34$~fs,  $\xi=0.2$, and $\varphi=\pi/2$. In excellent agreement with our theoretical predictions, the BW spectrum [Fig.~\ref{fig:2DSims}(a)] is very narrow compared to the FW one [Fig.~\ref{fig:2DSims}(b)].
Moreover, estimating the length $L$ of the plasma channel to be $30~\mu$m [Fig.~\ref{fig:2DSims}(d)] and evaluating Eq.~(\ref{eqn:Spectshape}) leads to a BW spectral $1/e$-width of $\Delta\nu=5$~THz, which coincides with the numerical obtained one at $z=-60~\mu$m. In time domain, the ratio between on-axis BW and FW THz field amplitude in Fig.~\ref{fig:2DSims}(d) is about 0.1.
To further validate the interference to be the governing effect shaping the BW spectrum, we linearly propagate a two-color Gaussian pump pulse in 3D geometry and use Jefimenko's Eq.~(\ref{eq:jefimenko}) to compute the BW spectrum (3D Jefimenko approach). In order to get a comparable plasma channel (length and free electron density), we adjust initial width and amplitude to $w=25~\mu$m and $A=6$~GV/m compared to the previous 2D configuration.
The inset in Fig.~\ref{fig:2DSims}(a) shows the resulting BW spectrum, reproducing again the predicted as well as simulated spectral width of $5$~THz. 

\begin{figure}
\includegraphics[width=\columnwidth]{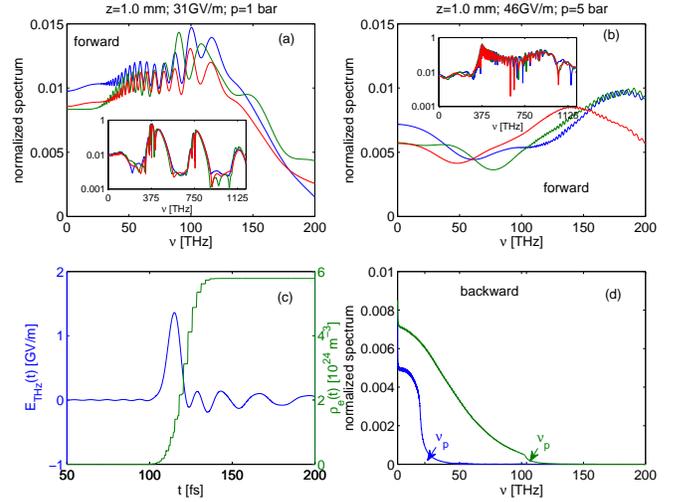}
 \caption{\label{fig:1DSims} 1D simulations: FW THz spectra for (a) $A=31$~GV/m at 1~bar and (b) $A=46$~GV/m at 5~bar from FDTD simulations (red line) and UPPE simulations with (green line) and without (blue line) third order nonlinear polarization included. The insets show the complete spectra. (c) illustrates that THz emission (green line) takes place mainly at the ionization front (blue line). (d) shows BW THz spectra corresponding to (a) (blue line) and (b) (green line) obtained from FDTD simulations. All spectra are normalized to $\lvert\hat{E}_\textrm{in}\left(\nu=375~\textrm{THz}\right)\rvert$.
}
\end{figure}

In the simple picture we have developed in
Fig.~\ref{fig:huiygensscheme} and Eqs.~(\ref{eq:jefimenko}) and (\ref{eqn:Spectshape}) we completely ignore the opacity of the generated plasma for frequencies below the plasma frequency $2\pi\nu_p=\sqrt{\rho_e q^2/ \epsilon_0 m_e}$. In the 2D examples above we find $\nu_p\approx45$~THz, but the generated plasma channel is small in transverse direction [$\approx 3~\mu$m, see Fig.~\ref{fig:2DSims}(d)] and THz radiation strongly diffracts, so we find our spectra still governed by the predicted interference effects. However, this may change if larger plasma volumes are created. In order to check this issue we resort to the extreme case of one-dimensional (1D) configuration, where we neglect all transverse spatial dependencies. Two-color pump pulses are launched in vacuum and hit a 1~mm thick argon layer at $z=0$~mm. This setup enables us to record BW spectra in FDTD simulations. For additional comparison, we simulate the FW fields by using a unidirectional pulse propagation equation (UPPE)~\cite{kolesik04}. This approach describes pump as well as the emitted THz fields in FW direction and was shown to reproduce experimental results~\cite{babushkin10a}. In our UPPE code, we can also include third order nonlinear polarization (Kerr effect), which is neglected in the FDTD simulations. It is worth noticing that the UPPE approach does not take into account the opacity of the plasma, whereas it is naturally included in the FDTD simulations. 
Figure~\ref{fig:1DSims} shows pump pulses with $\sigma_t=34$~fs,  $\xi=0.2$, and $\varphi=\pi/2$ and amplitudes  $A=31$~GV/m and $46$~GV/m launched into argon at pressure $p=1$~bar and $p=5$~bar, respectively. The plasma, generated in the $1$~mm long argon layer, is opaque for frequencies below $\nu_p=22$~THz ($A=31$~GV/m, $\rho_e^{\rm max}=6\times10^{24}$~m$^{-3}$) and $\nu_p=104$~THz ($A=46$~GV/m, $\rho_e^{\rm max}=13.5\times10^{25}$~m$^{-3}$). It turns out that for the FW emitted THz radiation the opacity of the plasma plays almost no role, we obtain excellent agreement between FDTD and UPPE simulations [see Fig.~\ref{fig:1DSims}(a,b)]. The reason for this effect is detailed in  Fig.~\ref{fig:1DSims}(c): the THz field is emitted before the plasma builds up, thus no damping influences its further FW propagation. In contrast, in BW direction plasma-opacity is important and only frequencies above the corresponding plasma frequencies can propagate through the already built up plasma. Therefore, in 1D geometry Eq.~(\ref{eqn:Spectshape}) is applicable only for frequencies above $\nu>\nu_p$, and for a plasma length of $1$~mm entirely destructive interference is expected. Indeed, this can be seen from the cutoffs of the BW spectra at the corresponding plasma frequencies in Fig.~\ref{fig:1DSims}(d). 
Below these frequencies $\nu_p$, radiation originates from the front-side of the plasma layer only, since contributions from deeper inside are damped.
 
In conclusion, we explained the differences in FW and BW emitted THz spectra from gas plasmas observed in full Maxwell numerical simulations by simple interference effects. Forward spectra are well described by unidirectional models and opacity of the plasma has negligible influence. We believe that the predicted sensitivity of the width of the BW THz spectrum on the length of the plasma channel could be tested experimentally in a straight-forward manner.

This work was performed using HPC resources at Rechenzentrum Garching 
and from GENCI-CCRT (Grant 20XX-x2010106003). I.~B. gratefully acknowledges financial support by the DFG.


\pagebreak

\end{document}